\documentclass[a4paper,11pt]{article}
\pdfoutput=1 % if your are submitting a pdflatex (i.e. if you have images in pdf, png or jpg format)

\usepackage{jheppub} % for details on the use of the package, please see the JHEP-author-manual
\usepackage[T1]{fontenc} % if needed
\usepackage[all]{hypcap} % solution to the problem that when you link to a float using hyperref, the link anchors to below the float's caption, rather than the beginning of the float.
\usepackage{slashed} % for Feynman slash notation
\usepackage{multirow} % more control on table columns/rows
\usepackage{booktabs} % for better looking tables
\usepackage{parskip} % nice white line spacing between paragraphs
\usepackage[section]{placeins} % Prevents floats moving to another section

\graphicspath{ {../plots/} {plots/} } 
\def\slash#1{#1 \hskip-0.45em /}
\def\beq{\begin{equation}}
\def\eeq{\end{equation}}
\def\bea{\begin{eqnarray}}
\def\eea{\end{eqnarray}}
\def\beqn{\begin{eqnarray}}
\def\eeqn{\end{eqnarray}}

\newcommand{\newc}{\newcommand}
\newc{\new}[1]{{\color{red}#1}}
\newc{\scale}{0.4}

\newc{\pmea}{{\mathbf p}}
\newc{\pmiss}{\slash{p}}
\newc{\ptmiss}{\slash{p}_T}
\newc{\pya}{p_{Y}}
\newc{\pyb}{p_{Y'}}
\newc{\pna}{p_{N}}
\newc{\pnb}{p_{N'}}
\newc{\pxa}{p_{X}}
\newc{\pxb}{p_{X'}}
\newc{\vlam}{{\bf \Lambda}}
\newc{\lamx}{\lambda_{\tmx^2}}
\newc{\lamn}{\lambda_{\tmn^2}}
\newc{\lamya}{\lambda_{\pya^2}}
\newc{\lamyb}{\lambda_{\pyb^2}}
\newc{\lammya}{\lambda_{\pmiss\pya}}
\newc{\lammyb}{\lambda_{\pmiss\pyb}}
\newc{\lamm}{\lambda_{\pmiss^2}}
\newc{\lamd}{\lambda_{\Delta}}
\newc{\lamP}{\lambda_{P^2}}

\newc{\amm}{\alpha}
\newc{\mya}{\beta}
\newc{\myb}{\beta'}
\newc{\yya}{\epsilon}
\newc{\yyb}{\epsilon'}

\newc{\mx}{m_{X}}
\newc{\mn}{m_{N}}
\newc{\my}{m_{Y}}

\newc{\tm}{\tilde{\bf {m}}}
\newc{\tmmax}{\tilde{\bf {m}}^{\rm max}}
\newc{\truem}{{\bf m}^{\rm true}}
\newc{\tmx}{\tilde{m}_{X}}
\newc{\tmn}{\tilde{m}_{N}}

\newc{\pp}{(p_\ell \cdot p_{\ell'} )}
\newc{\tmnsqmax}{(\tmn^{\rm max})^2}
\newc{\tmxsqmax}{(\tmx^{\rm max})^2}
\newc{\tmnmax}{\tmn^{\rm max}}
\newc{\tmxmax}{\tmx^{\rm max}}
\newc{\tmxmin}{\tmx^{\rm min}}

\newc{\detM}{{\mathcal M}}

\title{\boldmath Sharpening $m_{T2}$ cusps:  the mass determination of semi-invisibly decaying particles from a resonance}

\author[a]{Lucian A. Harland-Lang,}
\author[b]{Chun-Hay Kom,}
\author[c]{Kazuki Sakurai,}
\author[d]{Marco Tonini}

\emailAdd{lucian.harland-lang@durham.ac.uk}
\emailAdd{kom@hep.phy.cam.ac.uk}
\emailAdd{kazuki.sakurai@kcl.ac.uk}
\emailAdd{marco.tonini@desy.de}

\affiliation[a]{Department of Physics and Institute for Particle Physics Phenomenology, \\University of Durham, DH1 3LE, UK}
\affiliation[b]{Department of Mathematical Sciences, University of Liverpool, Liverpool L69 3BX, UK}
\affiliation[c]{Department of Physics, Theoretical Particle Physics and Cosmology, \\King's College London, London WC2R 2LS, UK} 
\affiliation[d]{DESY Theory Group, Notkestr.~85, 22607 Hamburg, Germany \\ \vspace{5pt}}

\preprint{\parbox{4cm}{IPPP/13/103\\DCPT/13/206\\KCL-PH-TH/2013-48\\LCTS/1014-09\\DESY-13-258}}

\keywords{$m_{T2}$, Kinematic variables, Consistent mass regions, Invisible mass, Mass measurements, LHC Phenomenology, Supersymmetry.}

\arxivnumber{1312.5720}

\abstract{
We revisit mass determination techniques for the minimum symmetric event topology, namely $X$ pair production followed by $X \to \ell N$, where $X$ and $N$ are unknown particles with the masses to be measured, and $N$ is an invisible particle, concentrating on the case where $X$ is pair produced from a resonance. We consider separate scenarios, with different initial constraints on the invisible particle momenta, and present a systematic method to identify the kinematically allowed mass regions in the $(m_N, m_X)$ plane. These allowed regions exhibit a cusp structure at the true mass point, which is equivalent to the one observed in the $m_{T2}$ endpoints in certain cases. By considering the boundary of the allowed mass region we systematically define kinematical variables which can be used in measuring the unknown masses, and find a new expression for the $m_{T2}$ variable as well as its inverse.  We explicitly apply our method to the case that  $X$ is pair produced from a resonance, and as a case study, we consider the process  $pp \to A \to \tilde \chi_1^+ \tilde \chi_1^-$, followed by $\tilde \chi_1^\pm \to \ell^{\pm} \, \tilde \nu_{\ell}$, in the Minimal Supersymmetric Standard Model and show that our method provides a precise measurement of the chargino and sneutrino masses, $m_X$ and $m_N$, at 14\,$\mathrm{TeV}$ LHC with $300 \, \mathrm{fb}^{-1}$ luminosity.
}

\begin{document}

%%%--- Title Page ---%%%
\maketitle
\flushbottom

%%%--- Introduction ---%%%
\section{Introduction}
\label{sec:introduction}
The new physics search program at the Large Hadron Collider (LHC) is soon to enter its second phase. If new physics is observed at the LHC, the masses of the Beyond the Standard Model (BSM) particles will be one of the first observables to be measured. The strategy for measuring the masses of these particles is in general strongly dependent on the event topology but, interestingly, one particular case is predicted in a range of BSM models: the pair production of BSM particles, each of which subsequently decays, through cascade decay chains, to an invisible particle. So far, most studies have focused on relatively long (2 -- 4 steps) 2--body cascade chains or short 3--body decay chains, initiated by the production of coloured BSM particles~\cite{Hinchliffe:1996iu, Hinchliffe:1998ys, Allanach:2000kt, Desch:2003vw, Kawagoe:2004rz, Cheng:2007xv, Cheng:2008mg, Cheng:2009fw, Webber:2009vm, Nojiri:2010dk, Burns:2008va, Konar:2009qr, Cho:2007qv, Cho:2007dh, Nojiri:2008hy, Nojiri:2008vq, Kim:2009nq, Nojiri:2010mk, Pietsch:2012nu}\footnote{See also~\cite{Barr:2010zj} for a review.}. However, the mass of coloured BSM particles is now strongly constrained by the null results of the BSM searches at the LHC. In the context of the Minimal Supersymmetric Standard Model (MSSM), the observation of a Higgs--like particle with $m_H \simeq 126$ GeV may indicate that squarks are heavier than the LHC reach~\cite{Giudice:2011cg, Ibe:2011aa}.

\begin{figure}[!ht]
	\begin{center}
		\includegraphics[scale=\scale]{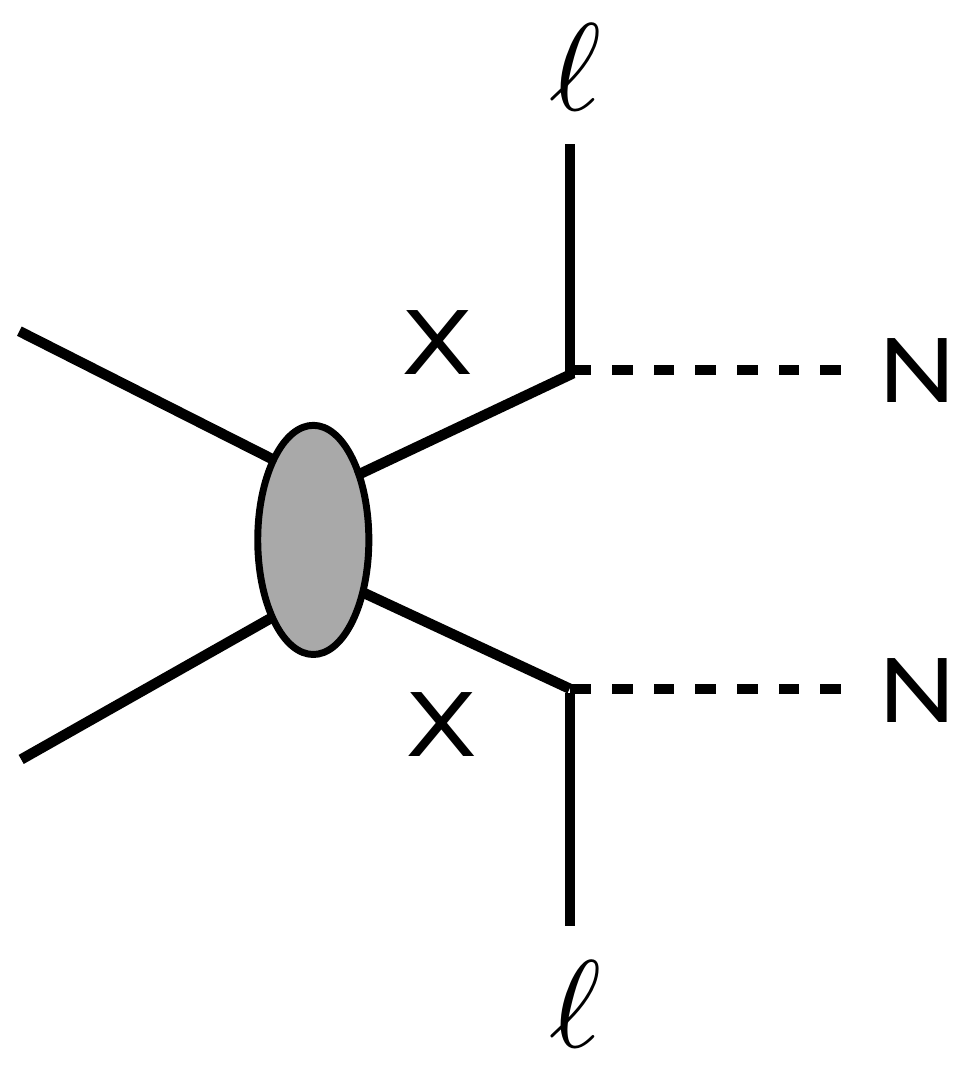}
	\caption{One--step decay chain of a pair produced and semi--invisibly decaying particle $X$.}
	\label{fig:diagram_1stepdecay}
	\end{center}
\end{figure}

On the other hand, constraints on colour--singlet BSM particles are much weaker. However, as the decay chain is in this case typically a short one--step process, namely $X$ pair production followed by $X \to \ell N$ (see figure~\ref{fig:diagram_1stepdecay}), where $N$ is an invisible particle, measuring the two masses $m_{N}$ and $m_{X}$ is particularly challenging.

At a hadron collider, this event topology yields the ``minimal'' set of constraints
\begin{eqnarray}
	\Phi_{\rm min}: 
		\left\{ \begin{array}{ll}
			\tmx^2 = (p_{\ell_1}^{\mu} + p_{N_1}^{\mu})^2  = (p_{\ell_2}^{\mu} + p_{N_2}^{\mu})^2  
			\vspace{2mm} \\
			\tmn^2 = p_{N_1}^2  = p_{N_2}^2 
			\vspace{2mm} \\
			\pmiss^T = {\bf p}_{N_1}^T + {\bf p}_{N_2}^T 
		\end{array} \right.
	\label{eq:phimin}
\end{eqnarray}
where ($\tmn, \tmx$) need not coincide with the true mass values $\truem\equiv(\mn,\mx)$, as they are a priori unknown. This set of constraints restricts the possible values of $\tmn$ and $\tmx$ and identifies  a kinematically allowed region in the ($\tmn, \tmx$) plane on an event--by--event basis. Furthermore, it is known~\cite{Cheng:2008hk} that the boundary of this allowed region under the $\Phi_{\rm min}$ constraints coincides with the $m_{T2}$ variable~\cite{Lester:1999tx}  
\begin{align}
	& \tilde{m}_{X; \Phi_{\rm min}}^{\rm min} (\tmn) = m_{T2}(\tmn) \equiv \label{eq:mt2} \\
	& \min_{ \sum_{i} {\bf p}_{N_i}^T = \pmiss^T } 
		\left\{
	\max \left[  m_T(p_{\ell_1}^{\mu}, {\bf p}^T_{N_1}, \tmn ), m_T(p_{\ell_2}^{\mu}, {\bf p}^T_{N_2}, \tmn ) \right]
		\right\} \nonumber		
\end{align}
where $m_T$ is the transverse mass~\cite{Arnison:1983zy}. In particular, the region with $\tmx(\tmn) < m_{T2}(\tmn)$ is excluded in the zero width limit and for perfect detector resolution.

\begin{figure}[!ht]
	\begin{center}
		\includegraphics[scale=0.33]{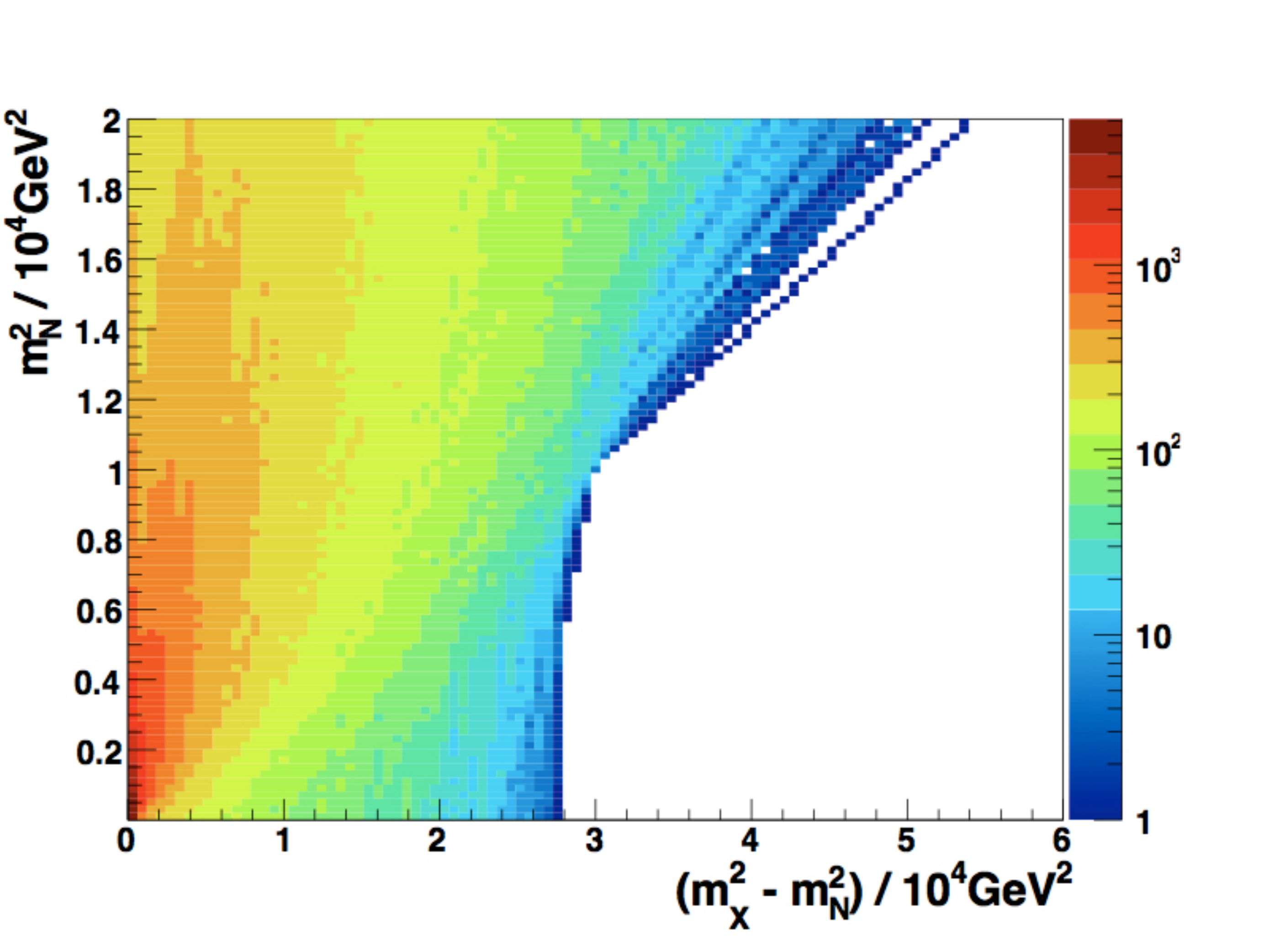}
		\caption{Density plot of $\Phi_{\rm min}$ boundary curves for $pp \to \tilde{q} \tilde{q}^{*} \to \tilde{\chi}^{+}_{1} \tilde{\chi}^{-}_{1} q \bar{q} \to \left( e^{+} \tilde{\nu}_{e} \right) \left( e^{-} \tilde{\bar{\nu}}_{e} \right) q \bar{q}$ LHC ($\sqrt{s}=14$ TeV) events with $(m_{\tilde q}, m_{\tilde{\chi}^{\pm}_{1}}, m_{\tilde{\nu}}) = (1500, 200, 100) \, \mathrm{GeV}$, at the generator level. The z--axis shows the number of boundary curves passing through $(0.06 \, \textrm{GeV}^2 ) \times (0.02 \, \textrm{GeV}^2)$ bins in $10^4$ events.}
		\label{fig:Phi_min}
	\end{center}
\end{figure}

If the system is boosted in the transverse direction by e.g.~hard initial state radiation (ISR), a collection of these $\tmx(\tmn)$ boundary curves from a large number of events exhibits a cusp structure~\cite{Cho:2007qv,Cho:2007dh,Barr:2007hy,Cheng:2008hk}. Figure \ref{fig:Phi_min} shows the density of the boundary curves projected onto the ($\tmx^2 - \tmn^2, \tmn^2$) plane, for the process $pp \to \tilde q \tilde q^{*}$, $\tilde q \to q \tilde{\chi}^{+}_{1}$, $\tilde{\chi}^{+}_{1} \to \ell^{+} \tilde{\nu}_{\ell}$, with $(m_{\tilde q}, m_{\tilde{\chi}^{\pm}_{1}}, m_{\tilde{\nu}}) = (1500, 200, 100) \, \mathrm{GeV}$, and neglecting finite width effects and detector resolution. The combination of all the event--by--event kinematically allowed regions provides a ``global'' allowed region, corresponding to the right hand side white region in figure~\ref{fig:Phi_min}. Indeed, we find that the decay of the heavy squarks provides a ``kick'' to the di--$X$ system, and a large boost in the transverse direction is achieved. Consequently, a cusp structure at the true mass point is observed. However, the population of the boundary curves around the cusp is very low and the cusp structure is not very distinct, even in this ideal case. In practice, the observation of this cusp is made even more difficult due to momentum mismeasurement and potential background contamination\footnote{For studies along these lines, see~\cite{Konar:2009wn, Cohen:2010wv}. }.

\begin{figure}[!ht]
	\begin{center}
		\includegraphics[scale=\scale]{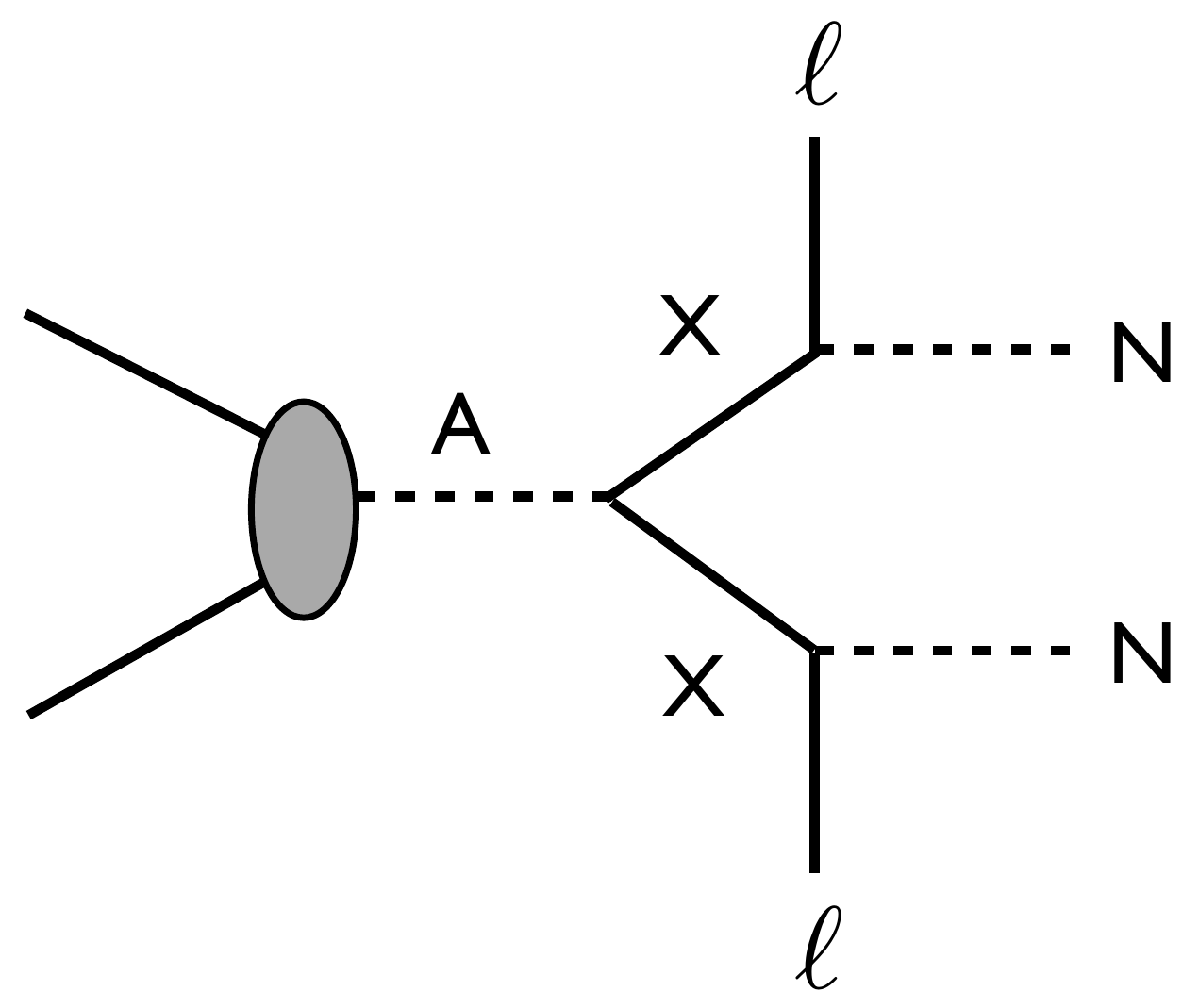}
		\caption{Di--$X$ production from a resonance $A$, followed by semi--invisible decays.}
		\label{fig:diagram_Adecay}
	\end{center}
\end{figure}

If one adds extra constraints to $\Phi_{\rm min}$, the kinematically allowed mass region is further restricted. Since the true mass point $\truem$ sits on the boundary of the global allowed region, adding such constraints will sharpen the cusp structure, and may make a simultaneous ($m_N, m_X$) measurement possible. A minimum and interesting possibility to extend $\Phi_{\rm min}$ is to add the constraint
\begin{equation}
	\Phi_{\rm s}: ~m_A^2 = (p_{\ell_1}^{\mu} + p_{N_1}^{\mu} + p_{\ell_2}^{\mu} + p_{N_2}^{\mu})^2 \; ,
	\label{eq:phis}
\end{equation}
which is relevant to the case that the particle $X$ is pair produced in the decay of a known resonance $A$ (see figure~\ref{fig:diagram_Adecay}).

One of the goals of this paper is to develop a method to extract $m_N$ and $m_X$ from event samples with the topology shown in figure~\ref{fig:diagram_Adecay}. As a benchmark scenario, we will investigate the LHC process $pp \to A \to \tilde{\chi}^{+}_{1} \tilde{\chi}^{-}_{1} \to \left( \ell^{+} \tilde{\nu}_{\ell} \right) \left( \ell^{-} \tilde{\bar{\nu}}_{\ell} \right)$, where $A$ is the CP--odd Higgs boson of the MSSM, and demonstrate that one can measure $m_{\tilde{\chi}^{\pm}_{1}}$  and $m_{\tilde{\nu}}$ with good accuracy at 14 TeV LHC with 300 fb$^{-1}$ of integrated luminosity.

\begin{figure}[!ht]
	\begin{center}
		\includegraphics[scale=0.33]{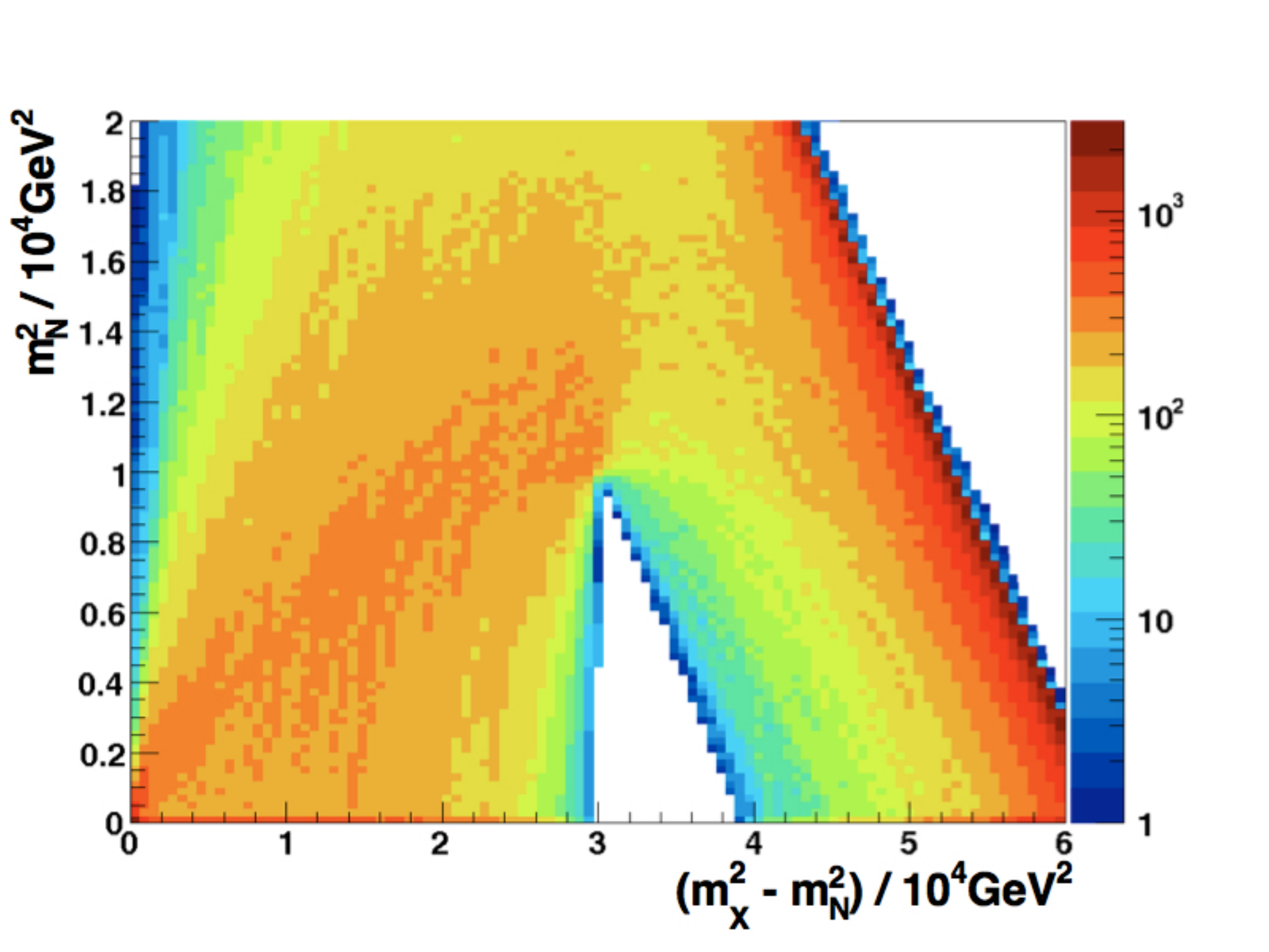}
		\caption{Density plot of the $\Phi_{\rm min} + \Phi_{\rm s}$ boundary curves for $pp \to A \to \tilde{\chi}^{+}_{1} \tilde{\chi}^{-}_{1} \to \left( \ell^{+} \tilde{\nu}_{\ell} \right) \left( \ell^{-} \tilde{\bar{\nu}}_{\ell} \right)$ LHC ($\sqrt{s}=14$ TeV) events with $(m_A, m_{\tilde{\chi}^{\pm}_{1}}, m_{\tilde{\nu}}) = (500, 200, 100) \, \mathrm{GeV}$, at the generator level. The z--axis shows the number of boundary curves passing through $(0.06 \, \textrm{GeV}^2 ) \times (0.02 \, \textrm{GeV}^2)$ bins in $10^4$ events.}
		\label{fig:Phi_min_Phi_S}
	\end{center}
\end{figure}

In figure~\ref{fig:Phi_min_Phi_S}, we show a density plot for the boundary curves of the event--by--event allowed mass regions for this process. For concreteness, we take $(m_A, m_{\tilde{\chi}^{\pm}_{1}}, m_{\tilde{\nu}}) = (500, 200, 100) \, \mathrm{GeV}$. One can see that the kinematically allowed region, given by the lower white triangle (we note that the allowed region for each event lies below the corresponding boundary curve, and so the upper white region is excluded) is more restricted with respect to the $\Phi_{\rm min}$ case of figure~\ref{fig:Phi_min} and that the cusp structure at the true mass point is more pronounced and more easily identified, reflecting the additional information which has been included, namely $\Phi_{\rm s}$.

Another way to extend $\Phi_{\rm min}$ is to assume that all four components of the missing momentum are known, namely by adding the constraint 
\begin{equation}
	\Phi_{\rm z}: ~\pmiss^z =  p^z_{N_1} + p^z_{N_2} \ .
\end{equation}
Notice that $\Phi_{\rm min} + \Phi_{\rm s} + \Phi_{\rm z} \equiv \Phi_{\rm max}$ is equivalent to $\Phi_{\rm min}$ with the last condition promoted to the Lorentz four--vector level $\pmiss^\mu = p_{N_1}^\mu + p_{N_2}^\mu$\,.

\begin{figure}[!ht]
	\begin{center}
		\includegraphics[scale=0.33]{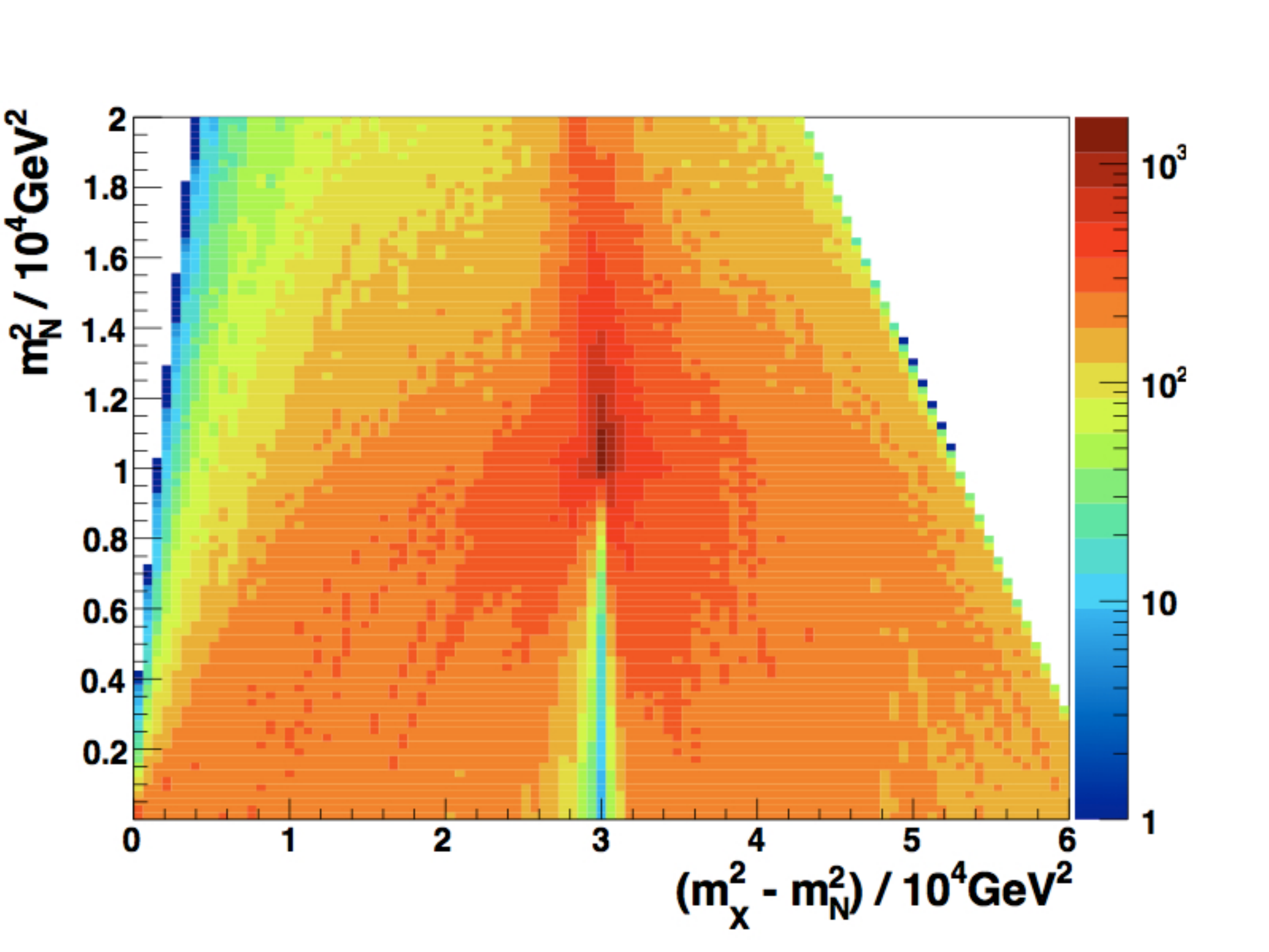}
			\caption{Density plot of $\Phi_{\rm min} + \Phi_{\rm s} + \Phi_{\rm z} \equiv \Phi_{\rm max}$ boundary curves of $e^{+} e^{-} \to \tilde{e}^{+} \tilde{e}^{-} \to \left( e^{+} \tilde{\chi}^{0}_{1} \right) \left( e^{-} \tilde{\chi}^{0}_{1} \right)$ ILC events with $(\sqrt{s}, m_{\tilde{e}}, m_{\tilde{\chi}^{0}_{1}}) = (500, 200, 100) \, \mathrm{GeV}$, at the generator level. The z--axis shows the number of boundary curves passing through $(0.06 \, \textrm{GeV}^2 ) \times (0.02 \, \textrm{GeV}^2)$ bins in $10^4$ events.}
			\label{fig:Phi_max}
	\end{center}
\end{figure}

This situation would be realised in a central exclusive process (CEP) with forward proton tagging at the LHC, $pp \to XX + \, pp$, $X \to \ell N$, or in the case of lepton colliders; a technique for extracting the masses ($m_{N},m_{X}$) in these cases has been studied previously in~\cite{HarlandLang:2011ih,HarlandLang:2012gn}. Notice that while at a lepton collider the invariant mass of the studied process is fixed by the center of mass energy of the collision, in the CEP case it is not a priori fixed, but rather is directly measured via proton tagging detectors. Assuming the set of constraints $\Phi_{\rm max}$, the global allowed region reduces to a straight line between the true mass point ($m^2_X-m^2_N, m_{N}$) and ($m^2_X-m^2_N, 0$), as can be seen in figure~\ref{fig:Phi_max}, allowing for a precise simultaneous ($m_{N}, m_{X}$) measurement. In figure~\ref{fig:Phi_max} we have shown equivalent density plots for a semi--invisible decay process at the ILC, namely $e^{+} e^{-} \to \tilde{e}^{+} \tilde{e}^{-} \to \left( e^{+} \tilde{\chi}^{0}_{1} \right) \left( e^{-} \tilde{\chi}^{0}_{1} \right)$ with $(\sqrt{s}, m_{\tilde{e}}, m_{\tilde{\chi}^{0}_{1}}) = (500, 200, 100) \, \mathrm{GeV}$.

The rest of this paper is organized as follows. In the next section we will describe the kinematic variables used for the mass determination in our study. We will focus on their analytical form and their relation with other known kinematical variables such as $m_{T2}$. Furthermore we will clarify how their distribution for a large number of events could provide a simultaneous ($m_{N},m_{X}$) mass measurement in a model--independent way. In the Results section we will then apply our method to the specific case of chargino and LSP mass measurement in events where two charginos are pair produced from the decay of the CP--odd Higgs $A$. Finally, we will summarize our results in the conclusions.

%%%--- Mass determination method ---%%%
\section{Mass determination method}
\label{sec:massmethod}
The use of the $\Phi_{\rm min} + \Phi_{\rm s} + \Phi_{\rm z} \equiv \Phi_{\rm max}$ constraints to develop a mass determination method~\cite{HarlandLang:2011ih,HarlandLang:2012gn} serves as a starting point for our discussion on the implementation of the $\Phi_{\rm min} + \Phi_{\rm s}$ constraints. In particular, the purpose of the method described in~\cite{HarlandLang:2011ih,HarlandLang:2012gn} was to determine all possible mass hypotheses $\tm \equiv (\tmn,\tmx)$ consistent with the mass--shell constraints, and when all four components of $\pmiss^\mu$ are known. We will begin with a summary of the method applied in~\cite{HarlandLang:2011ih,HarlandLang:2012gn}, before considering the $\Phi_{\rm min} + \Phi_{\rm s}$ case.

In general, any $p_{N_{1}}^\mu$ and $p_{N_{2}}^\mu$ satisfying $\pmiss^\mu = p_{N_{1}}^\mu + p_{N_{2}}^\mu$ can be parametrised as
\begin{equation}
	p^\mu_{N_{1}/N_{2}} = \frac{1\mp a}{2} \, \pmiss^\mu 
	\pm \frac{b}{2} \, p^\mu_{\ell_{1}} 
	\mp\frac{c}{2} \, p^\mu_{\ell_{2}} 
	\pm d \, P^\mu\,, 
	\label{eq:momenta}\\
\end{equation}
where $a,b,c,d$ are dimensionless constants, and $P^\mu$ is a space--like vector defined by $P_{\mu}\equiv\epsilon_{\mu\nu\rho\sigma}\pmiss^{\nu} p_{\ell_{1}} ^{\rho} p_{\ell_{2}}^{\sigma}$. Clearly we have $p^\mu_{X_{1}/X_{2}} = p^\mu_{N_{1}/N_{2}} + p^\mu_{\ell_{1}/\ell_{2}}$. With this parametrisation, the remaining $\Phi_{\rm max}$ constraints are given by
\begin{equation}
	\tmx^2 = p^2_{X_{1}}=p^2_{X_{2}}  
	,~~~~
	\tmn^2 = p_{N_{1}}^2  = p_{N_{2}}^2 ,
	\label{eq:mass}
\end{equation}
where again $\tm$ are test mass values which need not coincide with the true masses $\truem$. For a given $\tm$, the above four mass--shell conditions uniquely determine the coefficients $a, b, c$ (see~\cite{HarlandLang:2012gn} for the explicit forms) and yield the equation
\begin{equation}
	\lambda_N = \frac{c_a}{4 {\cal M}} \lambda_\Delta^2 + \frac{c_b}{2 {\cal M}} \lambda_\Delta + \frac{c_c }{4 {\cal M}} + d^2 \lambda_P^2 \,,
	\label{eq:quad}
\end{equation}
where $\lambda_N \equiv \tilde m^2_N/(p_{\ell_{1}} \cdot p_{\ell_{2}})$, $\lambda_\Delta \equiv (\tilde m^2_X - \tilde m^2_N)/(p_{\ell_{1}} \cdot p_{\ell_{2}})$, and where $c_a, c_b, c_c$ and ${\cal M}$ are functions of $p_{\ell_{1}}, p_{\ell_{2}}$ and $\pmiss$~\cite{HarlandLang:2012gn}.

A hypothesis ${\tilde {\bf m}}$ is said to be consistent if the corresponding $\lambda_\Delta, \lambda_N$ lead to $d^2 > 0$, in order to obtain four--momenta $p_{i}^{\mu}$ with real components \eqref{eq:momenta}. In other words, in the ($\tmn, \tmx$) plane the region which leads to $d^{2}>0$ corresponds to kinematically consistent mass hypotheses, while the boundary of this region is identified from eq.~(\ref{eq:quad}) by setting $d = 0$. Furthermore, one can show that $c_a / 4 {\cal M} < 0$~\cite{HarlandLang:2012gn}, and thus the shape of the boundary is a parabola with negative curvature, containing the true mass point $\truem$ below its apex in the ($\tilde m^2_X - \tilde m^2_N,  \tilde m^2_N$) plane.  

\begin{figure}[!ht]
	\begin{center}
		\includegraphics[scale=0.7]{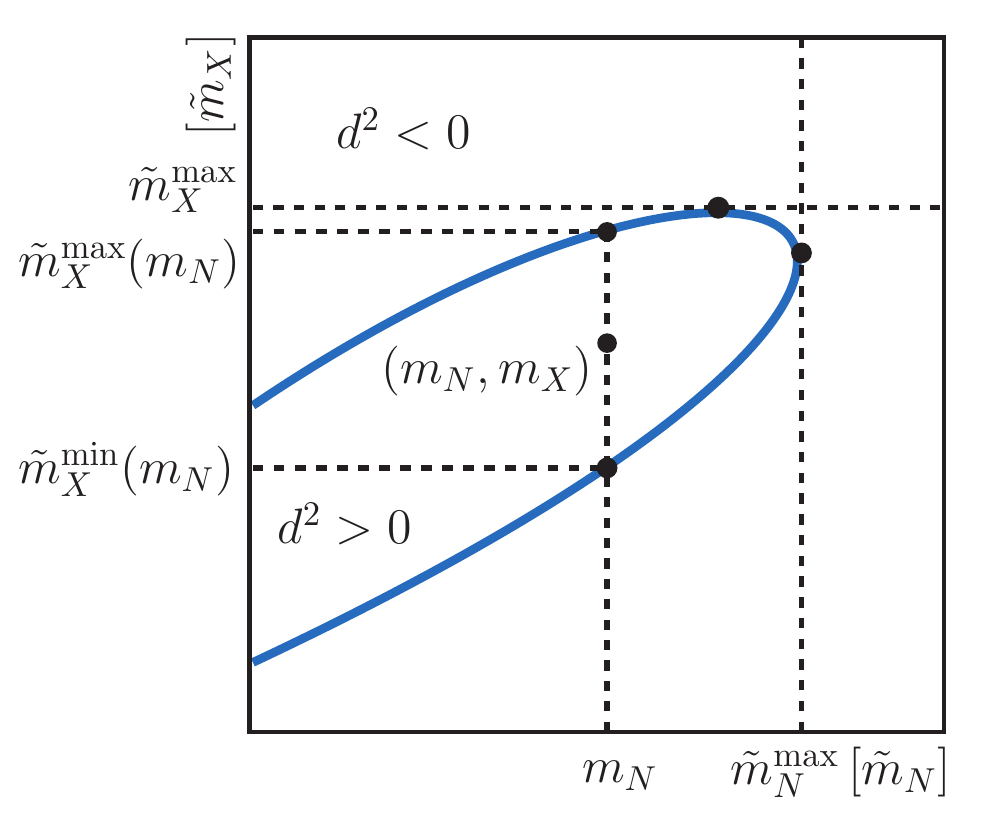}
		\caption{Kinematically consistent $(\tmn,\tmx)$ region ($d^2>0$ in eq.~\eqref{eq:quad}) for a ``typical'' event, defined by the four--momenta $(p_{\ell_{1}},p_{\ell_{2}},\pmiss)$, as published in \cite{HarlandLang:2012gn}.  The consistent mass region contains by definition the true mass point $(\mn,\mx)$. $\tilde{m}^{\rm max}_{N,X}$ is the maximum $\tmn$/$\tmx$ value, while $\tmx^{\rm min,max}(\mn)$ is the minimal/maximal value of $\tmx$ given $\mn$.}
		\label{fig:boundary}
	\end{center}
\end{figure}

As can be seen in figure~\ref{fig:Phi_min_Phi_S} and figure~\ref{fig:Phi_max}, the sharp cusp structure observed in the ($\tilde m^2_X - \tilde m^2_N,  \tilde m^2_N$) plane for the $\Phi_{\rm min} + \Phi_{\rm s}$ and $\Phi_{\rm max}$ cases would allow us to determine the true mass point by identifying the location of the cusps. Alternatively, one could define several {\it single} observables, whose distributions have endpoints at $m_X$ or $m_N$. Such observables would be more useful in handling background contamination, detector effects, experimental uncertainties and so on. We first define the global maximum of $\tmx$ and $\tmn$ along the boundary, which can be expressed analytically as~\cite{HarlandLang:2012gn}
\begin{align}
	(\tilde{m}_{X; \Phi_{\rm max}}^{\rm max})^2 &= \frac{p_{\ell_{1}} \cdot p_{\ell_{2}}}{4 {\cal M}} \Big[  c_c - \frac{(c_b + 2 {\cal M})^2}{c_a} \Big], \nonumber \\
	(\tilde{m}_{N; \Phi_{\rm max}}^{\rm max})^2 &= \frac{p_{\ell_{1}} \cdot p_{\ell_{2}}}{4 {\cal M}} \Big[  c_c - \frac{c_b^2}{c_a} \Big].
\end{align}
Other interesting variables which can be constructed are the extremal values of $\tmx$ along the boundary, for a given hypothesis on $\tmn$, denoted as $\tilde{m}_{X; \Phi_{\rm max}}^{\rm max/min}  (\tmn)$, and vice--versa $\tilde{m}_{N ; \Phi_{\rm max}} ^{\rm max/min}  (\tmx)$. Extending the results of~\cite{HarlandLang:2012gn}, we obtain their analytical form as
\begin{align}
	\big[ \tilde{m}_{X; \Phi_{\rm max}}^{\rm max/min}  (\tmn) \big]^2 &= \frac{p_{\ell_{1}} \cdot p_{\ell_{2}}}{c_a} \big[ C_X \pm \sqrt{D_X} \big], \nonumber \\
	\big[ \tilde{m}_{N ; \Phi_{\rm max}} ^{\rm max/min}  (\tmx) \big]^2 &= \frac{p_{\ell_{1}} \cdot p_{\ell_{2}}}{c_a} \big[ C_N \pm \sqrt{D_N} \big]
	\label{eq:m_m}
\end{align}
where
\begin{align}
	C_N &= c_a \lambda_X+ 2{\cal M} + c_b, \nonumber \\
	C_X &= c_a \lambda_N - c_b, \nonumber \\
	D_N &= (2{\cal M} + c_b)^2 + c_a ( 4 {\cal M} \lambda_X - c_c), \nonumber \\
	D_X &= c_b^2 + c_a ( 4 {\cal M} \lambda_N - c_c)
\end{align}
with $\lambda_X \equiv \tilde m^2_X/(p_{\ell_{1}} \cdot p_{\ell_{2}})$. Note that the assignment of the $\pm$ in eq.~(\ref{eq:m_m}) to the maximum or minimum mass depends on the sign of $(p_{\ell_{1}} \cdot p_{\ell_{2}})/c_a$, which may be negative. If the corresponding solution to the minimum $\tmn$ mass squared is negative, then this minimum mass does not lie on the boundary curve, and we therefore have $\tilde{m}_{N ; \Phi_{\rm max}} ^{\rm min}  (\tmx) = 0$. The consistent mass region for a ``typical'' event, and the new kinematic variables which can be extracted are shown in figure~\ref{fig:boundary}.

By definition, the boundary variables defined above possess the following properties:   
\begin{align}
	\tilde{m}_{X; \Phi_{\rm max}}^{\rm max} &\ge \tilde{m}_{X; \Phi_{\rm max}}^{\rm max} (m_N) \ge m_X,\nonumber \\
	\tilde{m}_{X; \Phi_{\rm max}}^{\rm min} &\le \tilde{m}_{X; \Phi_{\rm max}}^{\rm min} (m_N) \le m_X\; ,
	\label{eq:massphimax}
\end{align}
with similar relations for $N$. This observation has been used in~\cite{HarlandLang:2011ih,HarlandLang:2012gn} to show that the distributions of $\tilde{m}_{N; \Phi_{\rm max}}^{\rm max}$ and $\tilde{m}_{X; \Phi_{\rm max}}^{\rm max}$ exhibit a sharp endpoint structure at the corresponding true masses in the case of CEP process and $e^+ e^-$ colliders, allowing for a precise simultaneous ($m_{N}$, $m_X$) measurement.

We will now consider the $\Phi_{\rm min} + \Phi_{\rm s}$ case. Here, the energy and longitudinal components of $\pmiss$ are unknown, reflecting the normal situation at the LHC, where a significant and unknown proportion of the energy of the incoming hadrons in each event escapes down the beam pipe, and therefore the longitudinal and energy components of the missing momentum are not determined. However, as the right hand side of eq.~(\ref{eq:m_m}) is a function  of these unknowns, $\pmiss^0$ and $\pmiss^z$,  the boundary curve for the $\Phi_{\rm min} + \Phi_{\rm s}$ case and the corresponding kinematic variables can now be obtained by scanning over $\pmiss^0$ and $\pmiss^z$ under the constraint $\Phi_{\rm s}$, that is
\begin{align}
	\tilde{m}^{\rm max}_{X}(\tmn) &= \max_{\{ \pmiss^0, \pmiss^z \};  \Phi_{\rm s}} \big[ \tilde{m}^{\rm max}_{X; \Phi_{\rm max}}  (\tmn)  \big], \nonumber \\
	\tilde{m}^{\rm min}_{X}(\tmn) &= \min_{ \{ \pmiss^0, \pmiss^z \};  \Phi_{\rm s} } \big[ \tilde{m}^{\rm min}_{X; \Phi_{\rm max}}  (\tmn)  \big],
	\label{eq:phistmx}
\end{align}
with similar expressions for the $N$ case. The global maximum variables can be obtained as
\begin{align}
	\tilde{m}^{\rm max}_{X} &= \max_{ \{ \pmiss^0, \pmiss^z \};  \Phi_{\rm s}} \big[ \tilde{m}^{\rm max}_{X ; \Phi_{\rm max}} \big], \nonumber \\
	\tilde{m}^{\rm max}_{N} &= \max_{ \{ \pmiss^0, \pmiss^z \};  \Phi_{\rm s}} \big[ \tilde{m}^{\rm max}_{N ; \Phi_{\rm max}} \big].
	\label{eq:phisglobal}
\end{align}
By definition, analogous relations to eq.~\eqref{eq:massphimax} are valid in this case
\begin{align}
	\tilde{m}_{X/N}^{\rm max} &\ge \tilde{m}_{X/N}^{\rm max} (m_{N/X}) \ge m_{X/N},\nonumber \\
	\tilde{m}_{X/N}^{\rm min} &\le \tilde{m}_{X/N}^{\rm min} (m_{N/X}) \le m_{X/N}.
\end{align}
We will see in the following section that the kinematic variables $\tmnmax$ and $\tmx^{\rm min}(\tmn)$ in fact possess the best discriminating power for a simultaneous ($m_{N}, m_{X}$) measurement in the $\Phi_{\rm min} + \Phi_{\rm s}$ case.

We now consider the relation of our kinematic variables to the $m_{T2}$ variable. In analogy with the $\Phi_{\rm min}$ case \eqref{eq:mt2}, the kinematically allowed region under the $\Phi_{\rm min} + \Phi_{\rm s}$ constraints is in general bounded by $m_{T2}$  
\begin{equation}
	\tmx^{\rm min} (\tmn) \geq m_{T2} (\tmn) \,,
\end{equation}
where the inequality reflects the fact that additional information is provided by the $\Phi_{\rm s}$ constraint, further restricting $\tmx(\tmn)$. At the true invisible mass $m_{N}$ the endpoint of the $m_{T2}(\tmn = m_{N})$ distribution coincides with the true mass $m_{X}$~\cite{Cho:2007qv,Cho:2007dh}. Therefore, to draw a comparison and a cross--check of our method, for each event we will also evaluate the variable $m_{T2}(\tmn)$, and study its distribution for a large number of events.

Finally, we briefly return to the $\Phi_{\rm min}$ case. The boundary of the allowed mass region can be obtained in the same way as discussed above for the $\Phi_{\rm min} + \Phi_{\rm s}$ case, namely by scanning over $\pmiss^0$ and $\pmiss^z$ and taking the maximum or minimum depending on the variables. In figure~\ref{fig:Phi_min}, one can see that the allowed region is opened to $\tmx \to \infty$, that is the variables $\tilde m_{X; \Phi_{\rm min}}^{\rm max}(\tmn)$ and $\tilde m_{N; \Phi_{\rm min}}^{\rm min}(\tmx)$ are not defined. Knowing that the boundary curve in the $\Phi_{\rm min}$ case is given by $m_{T2}(\tmn)$ \eqref{eq:mt2}, we arrive at a new expression of the $m_{T2}$ variable
\begin{equation}
	m_{T2}(\tilde m_N) = \min_{\{ \pmiss^0, \pmiss^z \}} \big[ m^{\rm min}_{X; \Phi_{\rm max}}  (\tilde m_N)  \big]\,.
	\label{eq:mt2new}
\end{equation}
In the same way, an expression for the inverse $m_{T2}$ function can be written down
\begin{equation}
	m^{-1}_{T2}(\tilde m_X) = \max_{\{ \pmiss^0, \pmiss^z \}} \big[ m^{\rm max}_{N; \Phi_{\rm max}}  (\tilde m_X)  \big]\,.
\end{equation}
This function has the following properties
\begin{align}
	m^{-1}_{T2}(m_X) &\ge m_N, \nonumber \\
	m^{-1}_{T2}( m_{T2}(\tilde m_N) ) &= \tilde m_N.
\end{align}

%%%--- Results ---%%%
\section{Results}
\label{sec:results}
To illustrate the features of our method we will consider the case of associated production of the MSSM CP--odd Higgs $A$ with two b--jets, with the Higgs subsequently decaying into two charginos. We will then consider the decay of each chargino into a lepton plus a sneutrino. A final state with two opposite sign leptons, missing transverse energy and two b--jets will be therefore the topology under investigation. 

\begin{figure}[!ht]
	\begin{center}
		\includegraphics[scale=0.24]{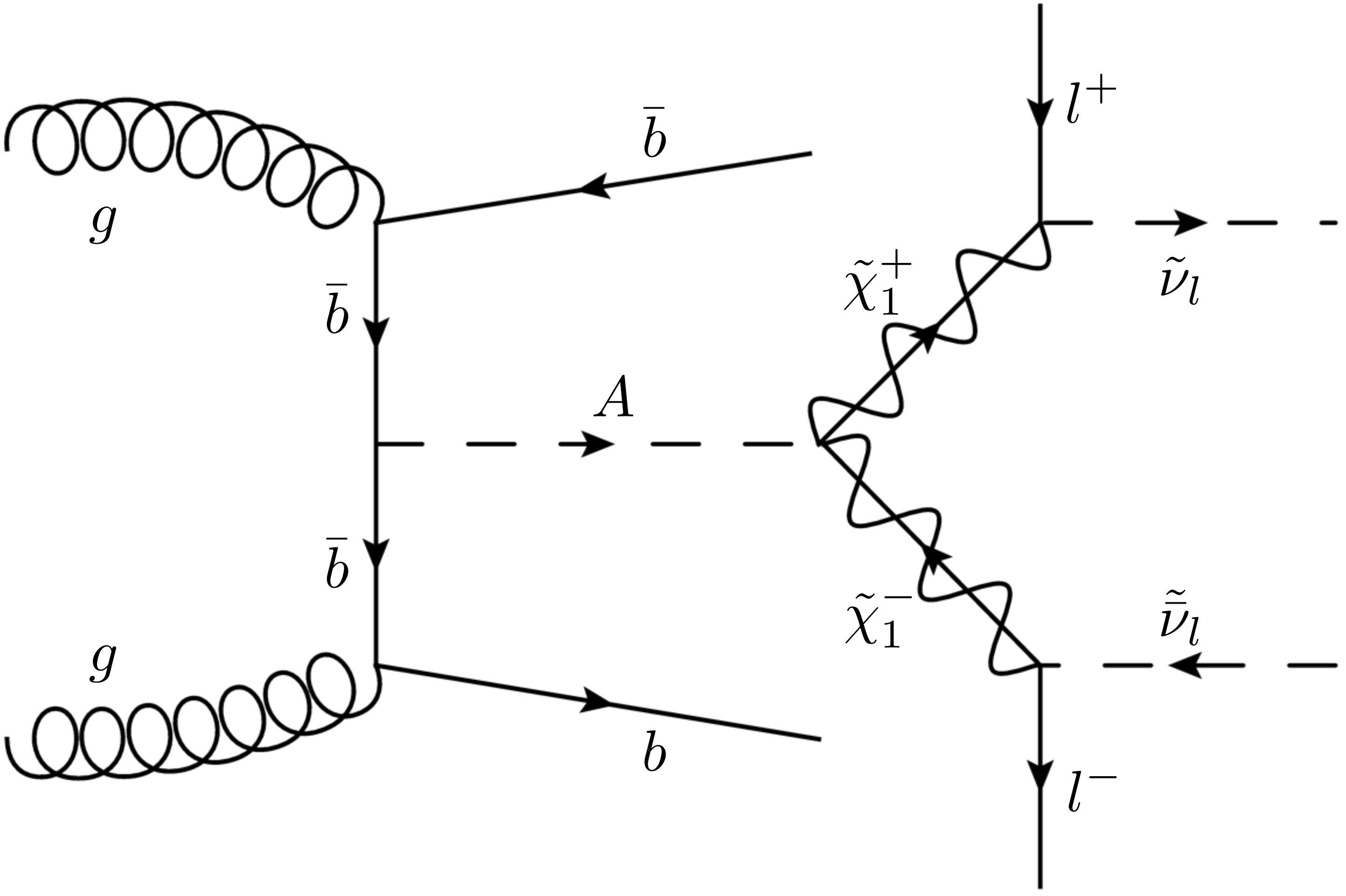}
		\caption{Feynman diagram for the $p \, p \to A b \bar{b} \to \tilde{\chi}_{1}^{+} \tilde{\chi}_{1}^{-} \, b \bar{b} \to \left( \ell^{+} \tilde{\nu}_{\ell} \right) \left( \ell^{-} \tilde{\bar{\nu}}_{\ell} \right) b \bar{b}$ process considered in our study.}
		\label{fig:Abb_diagram}
	\end{center}
\end{figure}

It is worth stressing that the method we have presented here is independent of the particular underlying model. However, for concreteness,  we have chosen a particular MSSM parameter space point, namely $m_{A} = 800 \, \textrm{GeV}$, $m_{\tilde{\chi}^{\pm}_{1}} = 350 \, \textrm{GeV}$ and $m_{\tilde{\nu}} = 200 \, \textrm{GeV}$.  We have chosen as reference values $\tan \beta = 50$ and $\mu = 400 \, \textrm{GeV}$, the former to increase the production cross section of the CP--odd Higgs for a given mass, the latter to increase the branching ratio into two charginos. Furthermore, $M2$ is set to $410 \, \textrm{GeV}$ in order to obtain the desired $m_{\tilde{\chi}^{\pm}_{1}}$, making the $\tilde{\chi}^{\pm}_{1}$ an admixture of Wino and Higgsino. Note that the dominant decay mode of the CP--odd Higgs is still into two bottom quarks, but we will assume that its mass $m_{A}$ has already been measured with  10$\%$ precision from a dedicated study of the $A \to \tau^{+} \tau^{-}$ channel, similar to~\cite{Aad:2012cfr}. We will however conclude this paper by showing how our method could also be used to obtain a quite precise determination of the CP--odd Higgs mass, without such input.

The dominant backgrounds for the considered final state consist of direct chargino pair production plus jets, and SM $t \bar{t}$ and $WW\textrm{+~jets}$ processes with the $W$ bosons decaying leptonically. A set of kinematic cuts has been chosen in order to maximize the signal over background ratio. In particular, each event is required to have exactly two opposite sign leptons with $|\eta| < 2.5$, and two b--jets with $p_{T}>20 \, \textrm{GeV}$, $|\eta| <2.5$. Moreover large cuts on the total missing transverse momentum ($\slashed{E}_{T} > 130 \, \textrm{GeV}$), the $p_{T}$ of the two leptons ($p_{T}^{\ell_{1}}>80 \, \textrm{GeV}$, $p_{T}^{\ell_{2}}>40 \, \textrm{GeV}$), and on the $m_{T2}$ variable ($m_{T2} >120 \, \textrm{GeV}$) are applied to successfully reduce the backgrounds. 

The associated CP--odd Higgs cross section has been calculated using FeynHiggs2.9.5~\cite{Hahn:2010te}. For the $\tilde{\chi}^{+}_{1} \tilde{\chi}^{-}_{1} \textrm{+~jets}$ process we used the LO cross section of chargino pair production with up to two matrix--element partons matched to the Pythia 6.42 parton shower via MLM merging scheme implemented in the MadGraph5--Pythia 6.42 interface~\cite{Alwall:2011uj,Sjostrand:2006za}. It is to be noted that the $\tilde{\chi}^{+}_{1} \tilde{\chi}^{-}_{1} \textrm{+~jets}$ cross section used in our simulation is not scaled by any NLO K-factor: however the contribution of this process is subdominant w.r.t.~the $t\bar{t}$ process, and we think that including higher order effects would not change our results. The values of the SM cross sections are reported in~\cite{Cacciari:2011hy,Campbell:2011bn}. The corresponding values are summarized in table~\ref{tab:crosssec}. 

We set the branching ratio of the chargino decay into charged lepton and sneutrino to $1.0$. We do not consider topologies in which the chargino decays into a $W$ boson and neutralino, and into a charged slepton and a neutrino, since they would be categorised in the $\Phi_{\rm min}$ case, the extra assumption $\Phi_{\rm s}$ being absent in these cases. In the MSSM the mass splitting between left--handed sneutrinos and left--handed charged sleptons is small, and if the phase space for $\tilde{\chi}^{\pm}_{1} \to \tilde{\ell}^{\pm} \, \nu_{\ell}$ is as large as $\tilde{\chi}^{\pm}_{1} \to \ell^{\pm} \, \tilde{\nu}_{\ell}$, the branching ratio for our target decay becomes about $0.5$. Sneutrinos can be significantly lighter than the left--handed charged sleptons if light right--handed sneutrinos are introduced together with a large $A$--term. In this case, the sneutrinos can be the only sfermions lighter than $\tilde{\chi}^{\pm}_{1}$ and $\textrm{BR} \left( \tilde{\chi}^{\pm}_{1} \to \ell^{\pm} \, \tilde{\nu}_{\ell} \right) = 1.0$ can be realised. It should be further noted that our procedure applies to both the cases in which the sneutrino is either long--lived or decays to invisible particles. The leptonic branching ratio of the $W$ boson is set to $0.216$~\cite{Beringer:1900zz}.

MadGraph5 is used to generate all parton--level events, which are then interfaced with the Pythia 6.42 parton shower. These are then passed to Delphes 3.0~\cite{deFavereau:2013fsa} to simulate the ATLAS detector in a fast manner, following the specifications reported in~\cite{Aad:2012cfr}. The public code described in~\cite{Cheng:2008hk} is used to evaluate $m_{T2}$ for each event.

\begin{table}[!ht]
	\centering
	\begin{tabular}{l | c c c c}
		\toprule[1pt]
		& $A b \bar{b}$ & $\tilde{\chi}^{+}_{1} \tilde{\chi}^{-}_{1} \textrm{+~jets}$ & $t \bar{t}$ & $WW\textrm{+~jets}$ \\
		\midrule[1pt]
		$\sigma \cdot \textrm{BR} \, [\textrm{pb}]$ & 0.023 & 0.079 & 40.92 & 5.80 \\
		\bottomrule[1pt]
	\end{tabular}
	\caption{Cross sections at LHC14 for the signal and background processes considered in our study, before cuts.}
	\label{tab:crosssec}
\end{table}

A signal over background ratio of $S/B \sim 6.5$ with roughly $1000$ remaining signal events is obtained with this setup, and the events passing the selection cuts are then used as input for our mass measurement method. In particular, we have simulated $100$ independent signal and background measurements at LHC14 with $300 \, \textrm{fb}^{-1}$ integrated luminosity, to evaluate a statistical uncertainty on our observables. 

In the following we attempt to determine $m_N$ by measuring the endpoint of the $\tmn^{\rm max}$ distribution: $m_{N}^{\rm exp}  \equiv (\tmn^{\rm max})^{\rm endpoint}$. We eventually determine $m_X$ by measuring the endpoint of the $\tmx^{\rm min}( m_N^{\rm exp} )$ distribution: $m_{X}^{\rm exp} \equiv [\tmx^{\rm min}(m_N^{\rm exp})]^{\rm endpoint}$. Because of detector resolution and a finite width of the decaying particles, the observed distributions might exceed the theoretical endpoints. Since off--shell effects of SUSY particles are negligible compared to the impact of detector resolution, we neglect the width effect for SUSY particles in our analysis. We have also neglected the $W$--boson width, which might as well cause a visible effect. Our estimation of the $WW$ and $t\bar{t}$ backgrounds are therefore possibly underestimated.

We have developed a numerical procedure to evaluate the endpoint of the different distributions, in order to minimise any potential bias in extracting these endpoints. In particular we randomly generate a large number of mass intervals, varying their midpoint and width, and store the ratio of events in the right--half of the interval over the number of events in the left--half of the interval (or left--half to right--half, depending on if it is a minimum or maximum endpoint being looked for, respectively). Each ratio is then weighted by the inverse of the interval width and by the number of events in the right--half (left--half) of the interval, so as to give greater weight to steeper drops and to more statistically significant drops, i.e. so that a drop from say $100$ to $50$ events receives a greater weight than a drop from $2$ to $1$ events. The distribution of the weighted ratios with respect to the corresponding midpoints should eventually peak at the position of the endpoint, which is then finally evaluated as the midpoint value with the highest weighted ratio.

As well as providing a measure of the endpoint position, this procedure allows us to determine an ``absolute'' value of the steepness of a distribution at its endpoint just by summing all the binned ratios: the simple idea is that the higher the sum, the ``steeper'' the endpoint. This steepness evaluation will be useful in the following.

\begin{figure}[!ht]
	\begin{center}
		\includegraphics[scale=\scale]{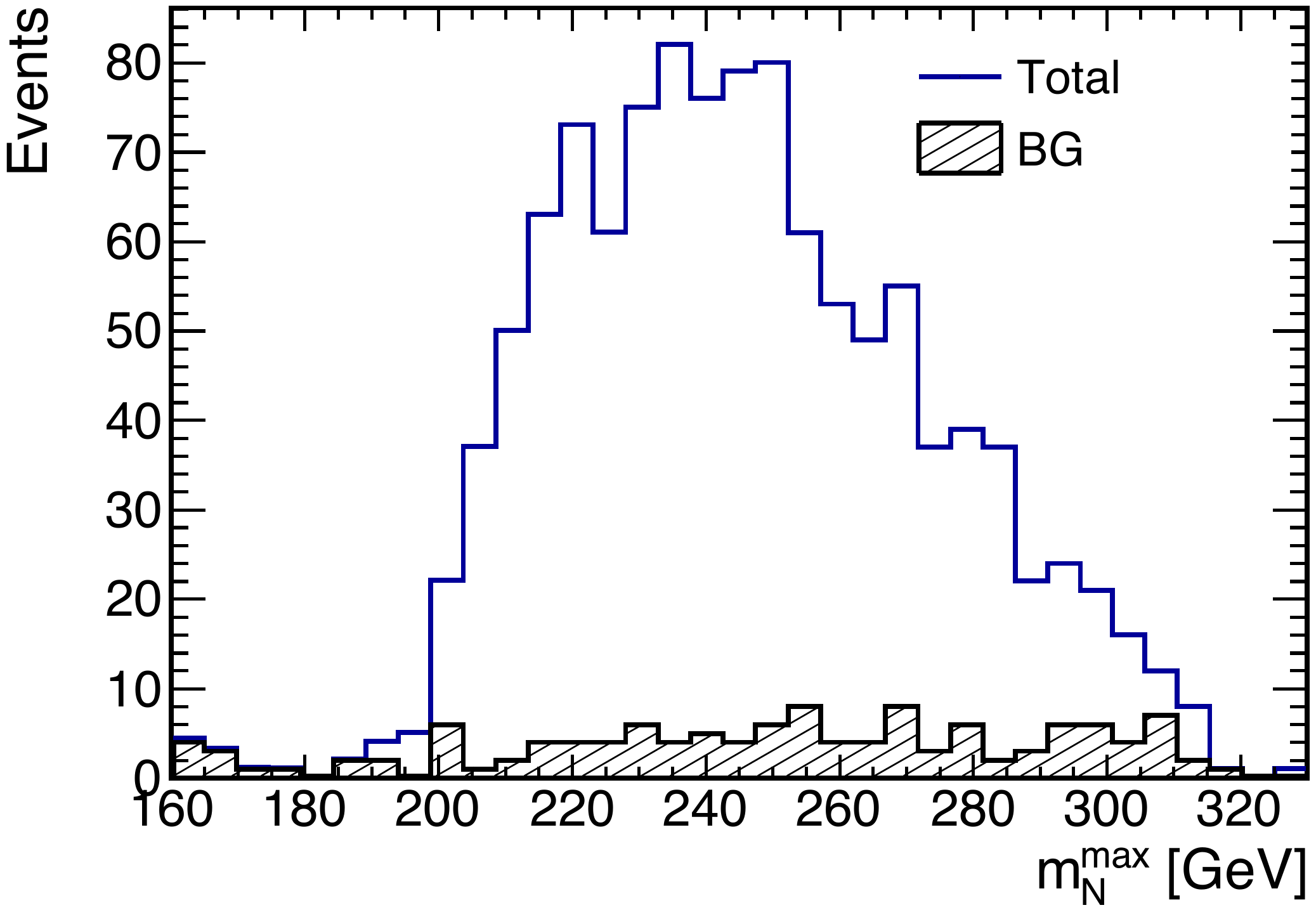}
		\caption{$\tmnmax$ distribution of a single signal and background simulation at LHC14 with $300 \, \textrm{fb}^{-1}$ integrated luminosity. We considered $m_{A} = 800 \, \textrm{GeV}$, $m_{\tilde{\chi}^{\pm}_{1}} = 350 \, \textrm{GeV}$ and $m_{\tilde{\nu}} = 200 \, \textrm{GeV}$.}
		\label{fig:800_200_mNmax}
	\end{center}
\end{figure}

In figure~\ref{fig:800_200_mNmax} we show a typical $\tmnmax$ distribution of a single LHC14 simulation. From our numerical procedure we can then evaluate the left--hand side endpoint, corresponding to the $m_{N}^{\rm exp}$ value of the single LHC14 simulation. By averaging over the 100 different simulations we obtain a measurement of the invisible mass $m_{N}^{\rm exp}$ of
\begin{equation}
	m_{N}^{\rm exp} = 195.9 \pm 2.5 \, \textrm{GeV}\;,
\end{equation}
remarkably close to the true value $m_{N} = 200 \, \textrm{GeV}$, where the uncertainty is calculated as standard deviation from the 100 independent measurements. 

\begin{figure}[!ht]
	\begin{center}
		\includegraphics[scale=\scale]{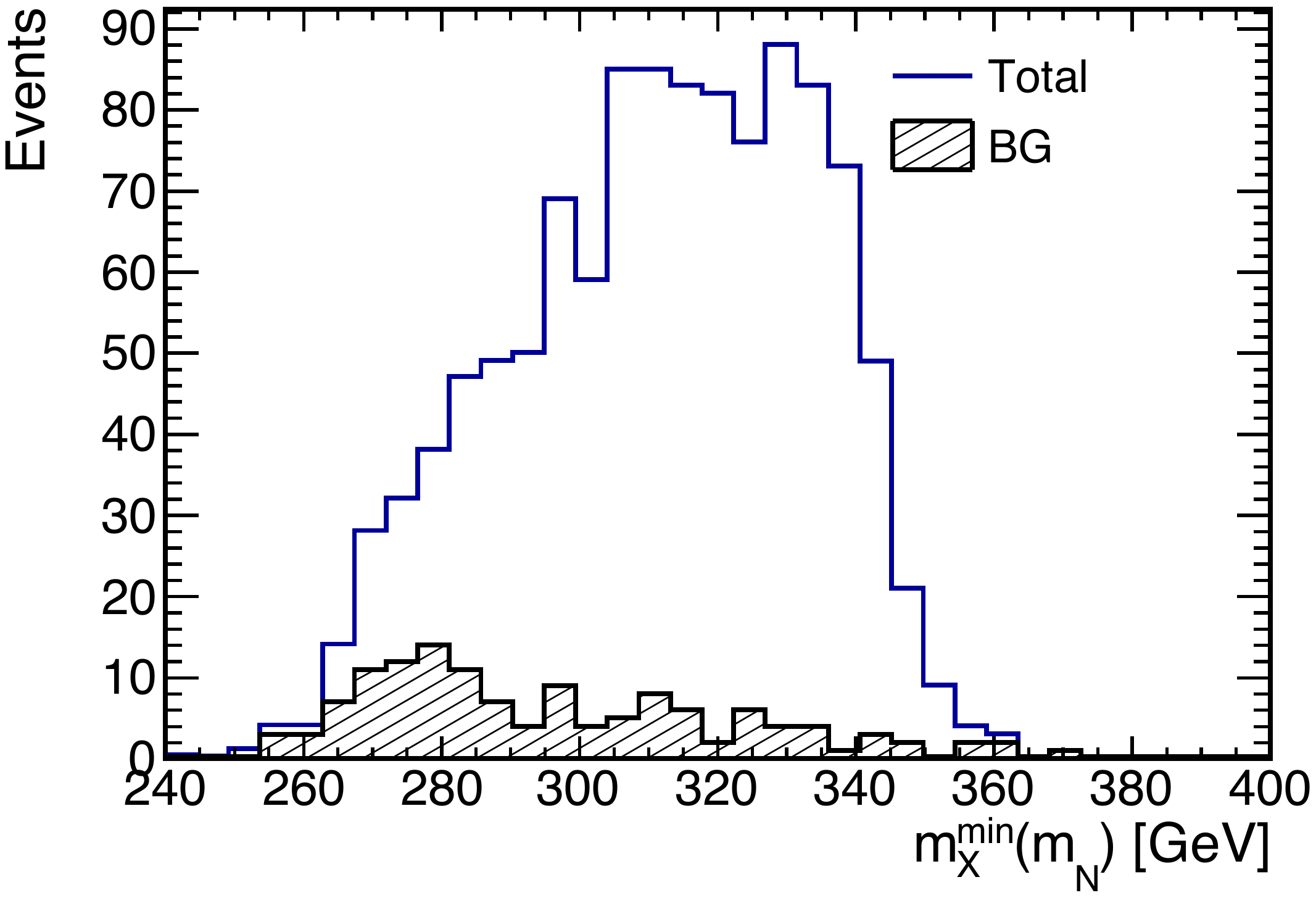}
		\caption{$\tmxmin (m_{N}^{\textrm{exp}})$ distribution of a single signal and background simulation at LHC14 with $300 \, \textrm{fb}^{-1}$ integrated luminosity. We considered $m_{A} = 800 \, \textrm{GeV}$, $m_{\tilde{\chi}^{\pm}_{1}} = 350 \, \textrm{GeV}$ and $m_{\tilde{\nu}} = 200 \, \textrm{GeV}$. The $m_{N}^{\textrm{exp}}$ value has been extracted from the endpoint of the $\tmnmax$ distribution.}
		\label{fig:800_196_mXmin}
	\end{center}
\end{figure}

In figure~\ref{fig:800_196_mXmin} we show a typical $\tmxmin (\tmn)$ distribution of a single LHC14 simulation, where $\tilde{m}_{N} = m_{N}^{\textrm{exp}} = 195.9 \, \textrm{GeV}$ is assumed. As before, we obtain a measurement of the chargino mass $m_{X}^{\rm exp}$ of
\begin{equation}
	m_{X}^{\rm exp} = 362.0 \pm 4.6 \, \textrm{GeV}\;,
\end{equation}
again close to the true value $m_{X} = 350 \, \textrm{GeV}$.

We can see in both cases that there is some difference between the true masses and those extracted from the endpoint measurements. This can be traced back to detector effects and to background contamination, which tend to smear the endpoints of the mass distributions, as can be seen in figures~\ref{fig:800_200_mNmax} and~\ref{fig:800_196_mXmin}, although with further refinements to our edge measurement technique, it is also possible that this offset may in general be reduced. However it is clear that this reasonably small effect can be corrected for in any experimental analysis, by comparing the measured endpoint values with simulation. 

\begin{figure}[!ht]
	\begin{center}
		\includegraphics[scale=\scale]{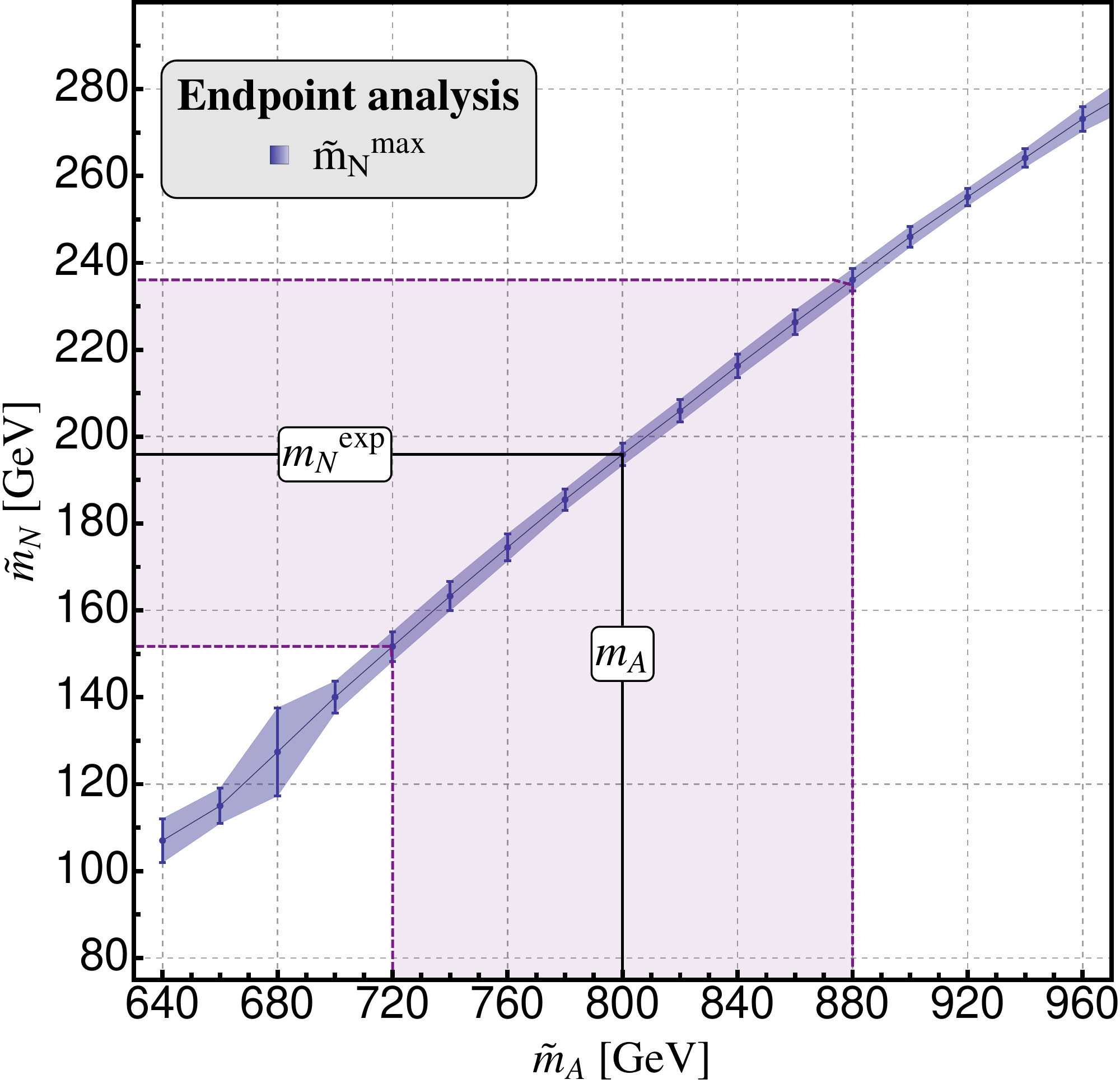}
		\caption{Endpoint measurements of the $\tmnmax (\tilde m_A)$ distribution for different  $\tilde{m}_{A}$ hypotheses. Each value represents the average endpoint measurement and corresponding standard deviation from the 100 independent LHC14 simulations. A band showing the effect of a $10\%$ uncertainty on $m_{A}$ is also shown.}
		\label{fig:mNmax_edge}
	\end{center}
\end{figure}

We have so far assumed that the mass of the resonance is precisely known. In realistic situations, our knowledge of $m_A$ is limited by the experimental uncertainty. To study this effect, we interpret the $m_A$ in eq.~\eqref{eq:phis} as a variable and allow the observables defined in eqs.~\eqref{eq:phistmx} and \eqref{eq:phisglobal} to depend on $\tilde m_A$, that is we have $\tilde m_{X}^{\rm min}(\tilde m_A, \tmn)$, $\tilde m_{N}^{\rm max}(\tilde m_A)$. In figure~\ref{fig:mNmax_edge} we plot the endpoints of the $\tmnmax (\tilde m_A)$ distribution for different hypotheses on $\tilde m_{A}$. It is also shown how a 10$\%$ uncertainty on $m_{A}$ affects this $m_{N}^{\rm exp}$ measurement, namely introducing a $\sim 20\%$ uncertainty. It is worth mentioning that the correlation among different test masses has been discussed in \cite{deGouvea:2012ez}.

\begin{figure}[!ht]
	\begin{center}
		\includegraphics[scale=\scale]{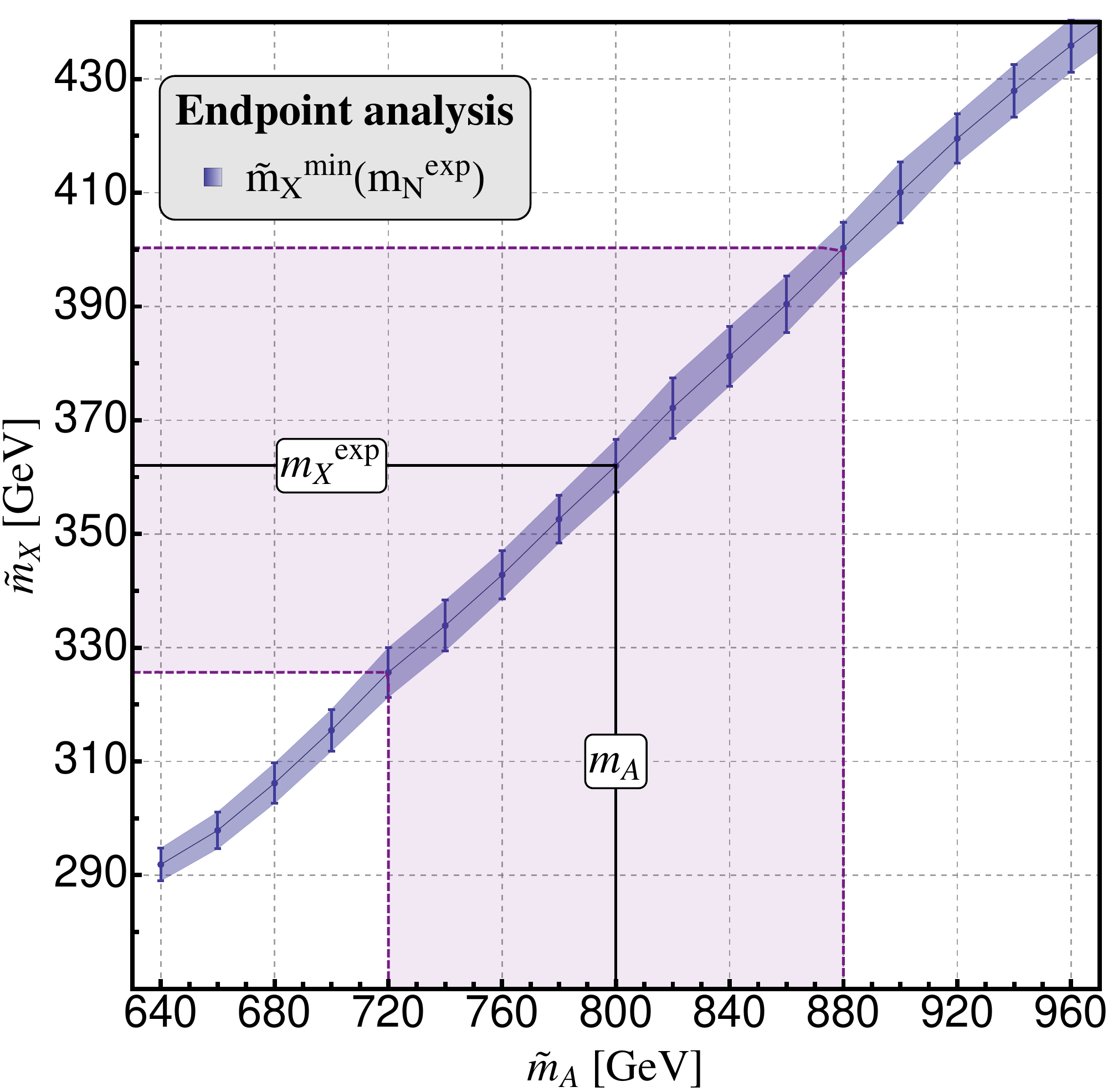}
		\caption{Endpoint measurements of the $\tmxmin (\tilde m_A, m_{N}^{\textrm{exp}})$ distribution for different  $\tilde{m}_{A}$ hypotheses. Each value represents the average endpoint measurement and corresponding standard deviation from the 100 independent LHC14 simulations. A band showing the effect of a $10\%$ uncertainty on $m_{A}$ is also shown.}
		\label{fig:mXmin_edge}
	\end{center}
\end{figure} 

The endpoints of the $\tmxmin (\tilde m_A, m_{N}^{\textrm{exp}})$ distribution are shown in figure~\ref{fig:mXmin_edge}: for each $\tilde{m}_{A}$ hypothesis, we have determined the corresponding $m_{N}^{\rm exp}$ value, and then used this as an input for the $\tmxmax (\tilde m_A, m_{N}^{\rm exp})$ distribution. It is also shown how a 10$\%$ uncertainty on $m_{A}$ affects the $m_{X}^{\rm exp}$ measurement, namely introducing a $\sim 20\%$ uncertainty.

\begin{figure}[!ht]
	\begin{center}
		\includegraphics[scale=\scale]{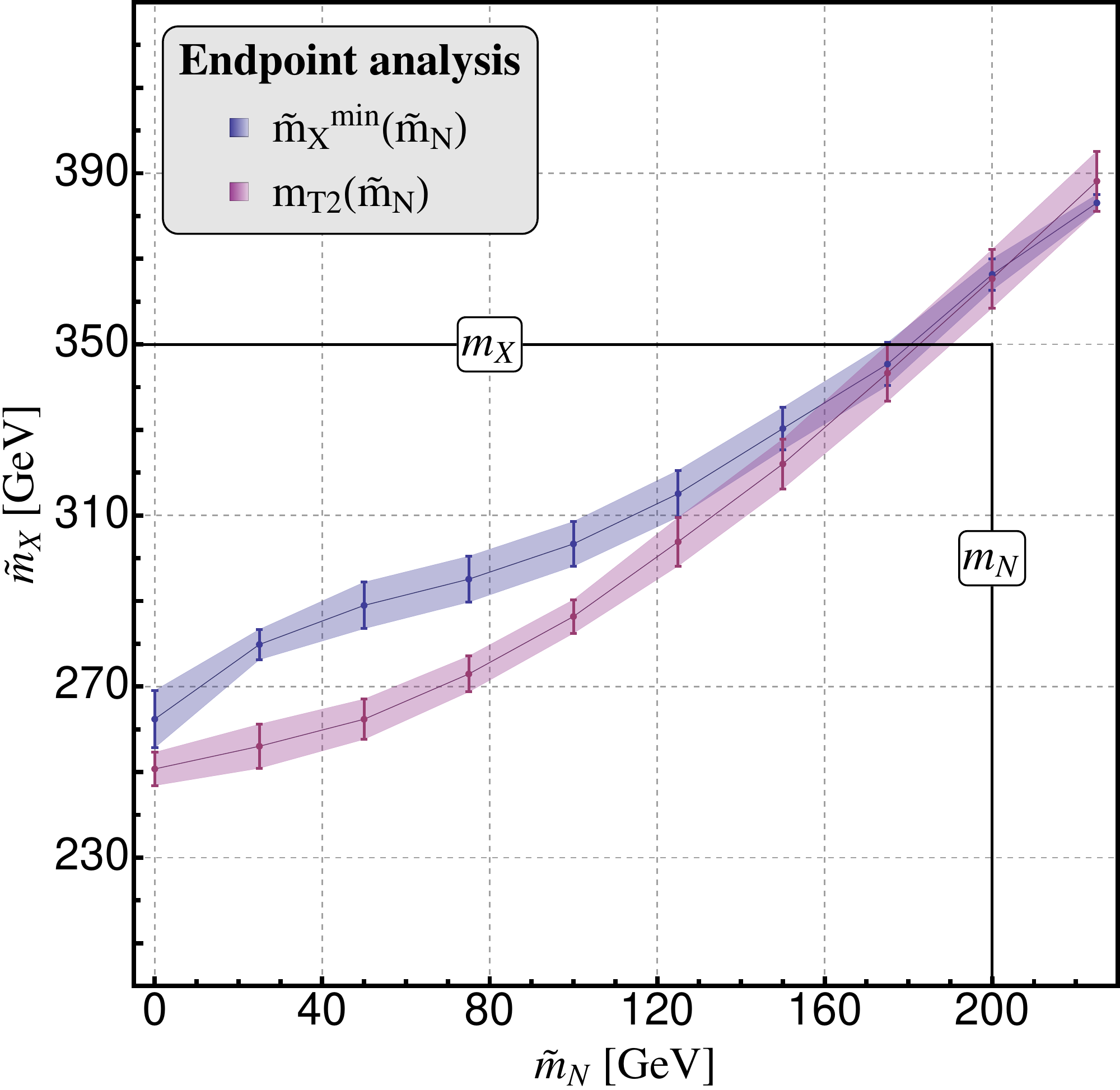}
		\caption{$\tmxmin(\tmn)$ and $m_{T2} (\tmn)$ distributions as functions of $\tmn$: they overlap in the vicinity of the true masses $\truem$.}
		\label{fig:endpoints}
	\end{center}
\end{figure}

It has previously been claimed that a simultaneous measurement of ($m_{N}, m_{X}$) is possible by using the kink structure arising in the distribution of the endpoints of the $m_{T2}(\tmn)$ variable as a function of $\tmn$, see~\cite{Cho:2007qv,Cho:2007dh,Cheng:2008hk}. However, this kink resides at the tail of the $m_{T2}(\tmn)$ distribution, making an accurate measurement difficult. On the other hand, even if such a kink structure is not evident, at the true invisible mass $m_{N}$ the endpoint of the $m_{T2}(\tmn = m_{N})$ distribution for a large number of events should always coincide with the \textit{mother} particle mass $m_{X}$, namely the chargino mass in our example. Therefore, by comparing the endpoints of the $\tmxmin (\tmn)$ and $m_{T2}(\tmn)$ distributions (assuming the true CP--odd Higgs mass $m_{A}$), we should be able to see that the two distributions coincide at $\truem$, as can be clearly seen from figure~\ref{fig:endpoints}, recalling that the endpoint measurement tends to overestimate by $\mathcal{O} (10 \, \textrm{GeV})$ the $m_{X}$ mass measurement.

The latter result should be viewed as a cross--check of the validity of our procedure rather than a direct measurement of the true masses, because of the rather large semi--overlapping region of the two curves. We can also see that there is not a clear kink structure in the $m_{T2}(\tmn)$ distribution, and thus this could not provide a precise mass measurement, at least for the case we have considered.

Throughout the previous sections, the mass of the resonance, $m_{A}$, has been assumed to be already (well) measured, to within $10\%$ uncertainty, in order to simultaneously evaluate ($m_{N}^{\textrm{exp}},m_{X}^{\textrm{exp}}$). However, if a wrong value for $m_A$ is used, then the $\Phi_{\rm s}$ constraint of (\ref{eq:phis}) no longer corresponds to the correct event kinematics, and one cannot expect the boundary variables, e.g. $\tmnmax$, to have a sharp endpoint structure. This observation may be used to measure the mass of the resonance.

\begin{figure}[!ht]
	\begin{center}
		\includegraphics[scale=\scale]{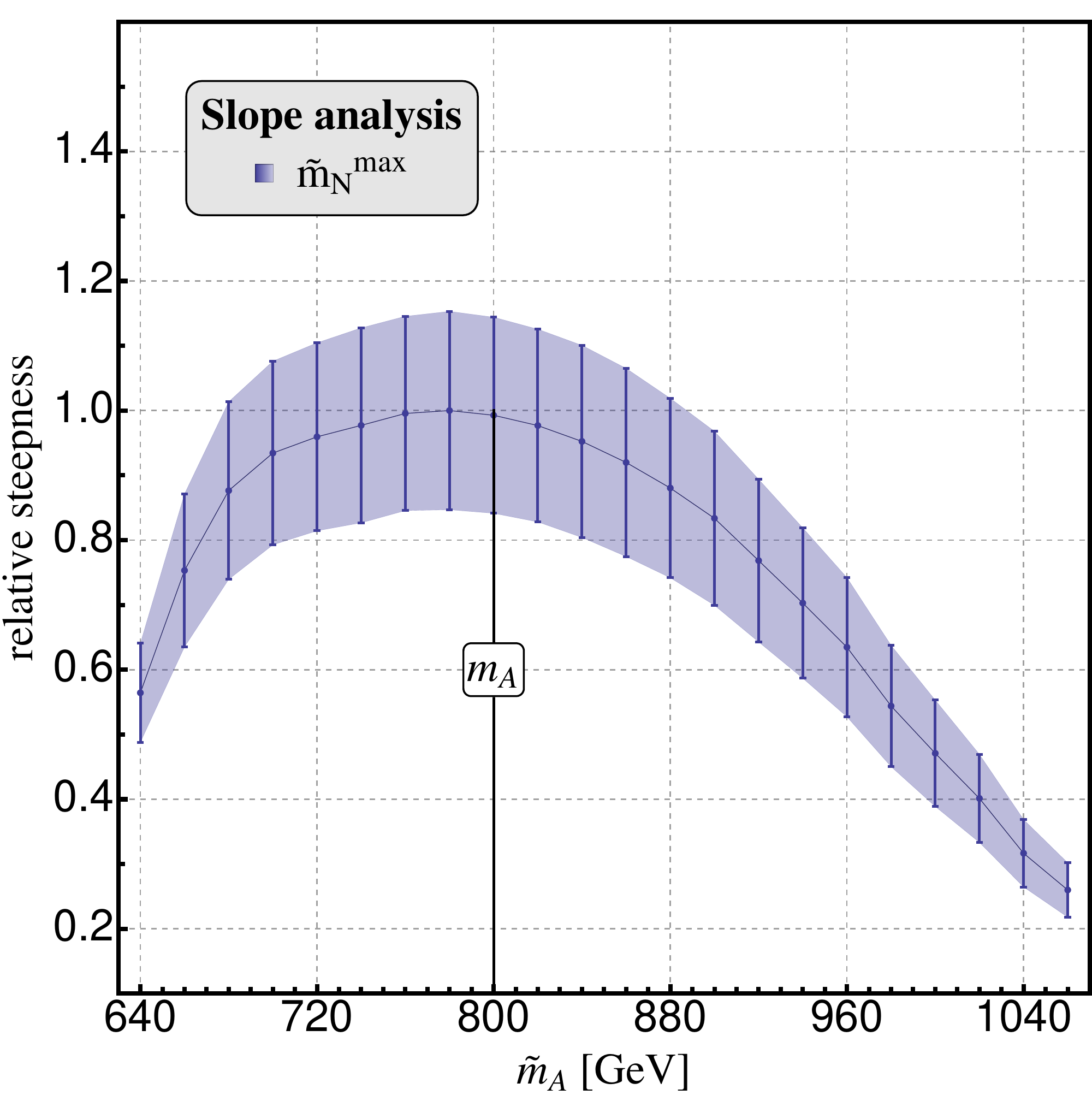}
		\caption{Slope measurement at the endpoint of the $\tmnmax (\tilde m_A)$ distribution: the values are normalised w.r.t.~the maximum measured steepness. Note that the maximum is observed near $\tilde{m}_{A} = m_{A}$.}
		\label{fig:mNmax_slope}
	\end{center}
\end{figure}

For example, one can expect the slope of $\tmnmax (\tilde m_A)$ at the endpoint to become steeper as the input value, $\tilde m_A$, approaches $m_A$, where we will expect a sharper endpoint structure. This feature is indeed seen in figure~\ref{fig:mNmax_slope}, where we plot the ``absolute steepness'' of the $\tmnmax$ distribution as a function of $\tilde m_A$, with the steepness evaluated from our numerical procedure described before. We have plotted the average values and corresponding standard deviations from the 100 independent LHC14 simulations, normalizing to the maximum steepness value for each simulation, such that the plot peaks at $1.0$. We expect this behavior to hold for other mass choices, but a more systematic understanding of this effect and its application to these and other mass measurements is the subject of ongoing studies. Furthermore, the steepness measurement is possible using also the $\tmxmin (\tilde m_A, m_{N}^{\rm exp})$ distribution, but we find a clearer peaking structure for the case of $\tmnmax (\tilde m_A)$ in our scenario.

Using this observation, we can eventually obtain a mass measurement of $m_{A}$, namely given by the $\tilde m_A$ hypothesis which provides the highest steepness of the $\tmnmax (\tilde m_A)$ distribution. By averaging over the 100 different LHC14 simulations, we finally measure $m_{A}^{\rm exp}$ as
\begin{equation}
	m_{A}^{\rm exp} = 776.4 \pm 34.3 \, \textrm{GeV}\;.
\end{equation}
The relatively large error on this value indicates how this result should be used only as a guide to infer the mass of the resonance $A$, although with further work on the precise manner in which the steepness of the $\tmnmax (\tilde m_A)$ distribution is evaluated, it may be possible to reduce this uncertainty.

%%%--- Conclusions ---%%%
\section{Conclusions}
\label{sec:conclusions}
In this work a model--independent method for mass measurements at hadron colliders, in semi--invisible decay chains of pair produced particles, has been discussed. We have considered as a benchmark the process $p \, p \to A b \bar{b} \to \tilde{\chi}_{1}^{+} \tilde{\chi}_{1}^{-} \, b \bar{b} \to \left( \ell^{+} \tilde{\nu}_{\ell} \right) \left( \ell^{-} \tilde{\bar{\nu}}_{\ell} \right) b \bar{b}$, where $A$ is the MSSM CP--odd Higgs. Here, the chargino $\tilde{\chi}_{1}^{\pm} \equiv X$ and LSP $\tilde{\nu}_{\ell} \equiv N$ masses are undetermined.  Analytic solutions of the final state system, taking into account the mass--shell conditions, constrain the possible ($\tmn,\tmx$) mass hypotheses consistent with the measured momenta for each event. Given this kinematically consistent mass region, one can then construct new useful variables, and the distribution of these from a large number of events is found to exhibit a sharp endpoint at the true chargino and LSP masses, respectively.

In particular we have shown that with this method one can obtain a precise measurement of ($\mn, \mx$) at the $\sqrt{s} = 14 \, \mathrm{TeV}$ LHC, with $300 \, \mathrm{fb}^{-1}$ of integrated luminosity. It is to be noted that the only additional information that has to be provided is the mass of the resonance $A$, from whose decay the charginos are pair produced. The total missing momentum is not required to be an input of our analysis, as was considered in~\cite{HarlandLang:2011ih,HarlandLang:2012gn}: our approach reflects a more common measurement scenario at the LHC.

Furthermore we have shown for our benchmark example that the value of the slope of the $\tmnmax$ distribution at the corresponding endpoint for different $\tilde m_{A}$ hypotheses develops a peak at the true mass $m_{A}$, and thus this fact could provide a guide to infer the mass of the resonance $A$. A more systematic understanding and application of this effect to mass measurements is the subject of ongoing studies.

%%%--- Acknowledgements ---%%%
\acknowledgments
We thank J\"urgen Reuter and Maikel de Vries for useful discussions. M.T. has been partially supported by the Deutsche Forschungsgemeinschaft within the Collaborative Research Center SFB 676 "Particles, Strings, Early Universe". K.S. has been partially supported by the London Centre for Terauniverse Studies (LCTS), using funding from the European Research Council via the Advanced Investigator Grant 267352.

%%%--- Bibliography ---%%%
\bibliographystyle{JHEP}
\bibliography{mt2kink}

%%%--- End Document ---%%%
\end{document}